\documentclass[showkeys,superscriptaddress,floatfix,prd,10pt,aps,twocolumn]{revtex4-2}
\usepackage{graphicx,epstopdf}
\usepackage[caption=false]{subfig}
\usepackage{appendix}
\usepackage[T1]{fontenc}
\usepackage{lmodern}
\usepackage[dvipsnames,x11names]{xcolor}
\usepackage[colorlinks=true,linkcolor=NavyBlue,citecolor=ForestGreen,urlcolor=NavyBlue]{hyperref}
\usepackage[sort&compress]{natbib}
\usepackage{lipsum}
\usepackage{morefloats}
\usepackage[pdf]{pstricks}
\usepackage{amsmath}
\usepackage{amssymb}
\usepackage{amsfonts}
\usepackage{rotating}
\usepackage{cancel}
\usepackage{mathtools}
\usepackage{bbm}
\usepackage{dsfont}
\usepackage{bbold}
\usepackage{multirow}
\usepackage{ulem}
\usepackage{physics}
\usepackage{orcidlink}
\usepackage{colortbl}

\def\xslash{x\!\!\!\slash }

\def\vel{\left|}
\def\ver{\right|}

\begin{document}

\title{Magnetic and quadrupole moments of the $Z_{c}(4020)^+$, $Z_{c}(4050)^+$, and  $Z_{c}(4600)^{+}$ states in the diquark-antidiquark picture}

\author{Ula\c{s}~\"{O}zdem\orcidlink{0000-0002-1907-2894}
}%
\email[]{ulasozdem@aydin.edu.tr }
\affiliation{Health Services Vocational School of Higher Education, Istanbul Aydin University, Sefakoy-Kucukcekmece, 34295 Istanbul, T\"{u}rkiye}

\date{\today}
 
\begin{abstract}
The magnetic and quadrupole moments of the $Z_{c}(4020)^+$, $Z_{c}(4050)^+$ and  $Z_{c}(4600)^{+}$ states are calculated within the QCD light-cone  sum rules. The compact diquark-antidiquark interpolating currents and the distribution amplitudes of the on-shell photon are used to extract the magnetic and quadrupole moments of these states. The magnetic moments are acquired as  $\mu_{Z_{c}} = 0.50 ^{+0.22}_{-0.22}~\mu_N$,  $\mu_{Z^{1}_{c}}=1.22 ^{+0.34}_{-0.32}~\mu_N$, and $\mu_{Z^2_{c}}=2.40 ^{+0.53}_{-0.48}~\mu_N$  for the  $Z_{c}(4020)^+$, $Z_{c}(4050)^+$ and  $Z_{c}(4600)^{+}$ states, respectively.  The magnetic moments evaluated for the  $Z_{c}4020)^+$, $Z_{c}(4050)^+$ and  $Z_{c}(4600)^{+}$ states are sufficiently large to be experimentally measurable. The magnetic moment is an excellent platform for studying  the internal structure of hadrons governed by the quark-gluon dynamics of QCD because it is the leading-order response of a bound system to a weak external magnetic field. The quadrupole moment results are  $\mathcal{D}_{Z_c}=(0.20 ^{+0.05}_{-0.04}) \times 10^{-3}~\mbox{fm}^2 $, $\mathcal{D}_{Z_c^1}=(0.57 ^{+0.07}_{-0.08}) \times 10^{-3}~\mbox{fm}^2 $, and $\mathcal{D}_{Z_c^2}=(0.30 ^{+0.05}_{-0.04}) \times 10^{-3}~\mbox{fm}^2 $ for the  $Z_{c}(4020)^+$, $Z_{c}(4050)^+$ and  $Z_{c}(4600)^{+}$ states, respectively. We obtain a non-zero, but small, value for the quadrupole moments of the $Z_c$ states, which indicates a non-spherical charge distribution. The nature and internal structure of these states can be elucidated by comparing future experimental data on the magnetic and quadrupole moments of the $Z_{c}(4020)^+$, $Z_{c}(4050)^+$, and $Z_{c}(4600)^{+}$ states with the results of the present study.
\end{abstract}

\maketitle

\section{Introduction}
Since the observation of the X(3872) state by the Belle collaboration in 2003 \cite{Belle:2003nnu}, many hadronic states have been observed that cannot be classified in the conventional two- or three-quark picture. The LHCb, BESIII, CDF, BABAR, CMS, Belle, and D0 collaborations have subsequently observed numerous states that cannot be classified in the conventional quark picture and are represented as XYZ particle, hybrid, and pentaquark states; however, some are still awaiting confirmation, and quantum numbers have not yet been assigned. The most important achievement in the field of non-conventional states has been the discovery of charged tetraquark states. Currently, there are ten members in the set of charged hidden-charmed tetraquark states: $Z_c(3900)$, $Z_c(4020)$, $Z_c(4050)$, $Z_c(4100)$,  $Z_c(4200)$, $Z_2(4250)$, $Z_c(4430)$,  $Z_c(4600)$, $Z_{cs}(3985)$, $Z_{cs}(4000)$, and $Z_{cs}(4220)$, reported in decays into final states that contain a pair of light and charm quarks~\cite{Choi:2007wga,Aaij:2014jqa,Mizuk:2008me,Ablikim:2013mio,Liu:2013dau,Ablikim:2013wzq,Ablikim:2013emm,Chilikin:2014bkk,Wang:2014hta,Collaboration:2011gja,LHCb:2019maw,BESIII:2020qkh,LHCb:2021uow}.  They cannot be classified as conventional charmonium mesons in because of their electric charge and must be nonconventional states with a minimum quark content $ c \bar {c} u \bar {d} $/$ c \bar {c} d \bar {u}$/$c \bar {c} s \bar {u} $/$ c \bar {c} u \bar {s}$.  The charged tetraquark states receive considerable  attention because of their different properties. The study of these states can help to not only  elucidate their nature and substructure but also obtain functional data on the nature of the strong interaction inside these particles. As tetraquark states with a $ c \bar {c} u \bar {d} $/$ c \bar {c} d \bar {u}$/$c \bar {c} s \bar {u} $/$ c \bar {c} u \bar {s}$ quark content, this family of charged states has generally been studied as molecular and compact diquark-antiquark pictures. Many comprehensive reviews on this subject can be found in literature~\cite{Faccini:2012pj,Esposito:2014rxa,Chen:2016qju,Ali:2017jda,Esposito:2016noz,Olsen:2017bmm,Lebed:2016hpi,Guo:2017jvc,Nielsen:2009uh,Brambilla:2019esw,Liu:2019zoy, Agaev:2020zad, Dong:2021juy,Chen:2015ata,Meng:2022ozq,Chen:2022asf}.

Magnetic and quadrupole moments are important parameters of hadrons that can be measured and calculated, just like mass and decay. In studying the internal structure and possible deformation of the hidden-charm tetraquark states, magnetic and quadrupole moments are of particular interest. The magnetic and quadrupole moments of hidden-charm tetraquark states have been extracted in several studies in the literature~\cite{Ozdem:2021yvo,Ozdem:2017jqh,Xu:2020qtg,Wang:2017dce,Xu:2020evn,Ozdem:2022kck,Ozdem:2021hka,Wang:2023vtx}. Moreover, in Ref. \cite{Li:2020rcg}, QCD calculations of radiative heavy-meson decay form factors were performed by including the subleading power corrections of the twist-two photon distribution amplitude in the next-to-leading order in $\alpha_s$ using QCD light-cone sum rules. In this study we evaluate the magnetic and quadrupole moments of the tetraquark states $Z_{c}(4020)^+$, $Z_{c}(4050)^+$, and $Z_{c}(4600)^{+}$ (hereafter $Z_{c}$, $Z^1_{c}$, and $Z^2_{c}$, respectively) by considering them as the diquark-antiquark picture within  QCD light-cone sum rules. The QCD light-cone sum rules method is a powerful technique for  studying exotic hadron properties and has been successfully applied to the calculation of masses, form factors, magnetic moments, decay constants, and so on. The correlation function is evaluated  in terms of both hadrons (the so-called hadronic part) and in terms of the quark-gluon degrees of freedom (the QCD part) according to the QCD light-cone sum rules technique.   By equating these two descriptions of the correlation function, the physical quantities, that is, the magnetic and quadrupole moments, are evaluated~\cite{Chernyak:1990ag, Braun:1988qv, Balitsky:1989ry}.

This paper is organized as follows.   In Sec. \ref{secII} we construct the QCD light-cone sum rules for the magnetic and quadrupole moments of the hidden-charm tetraquark states. The numerical results and the corresponding discussion of the magnetic and quadrupole moments of the compact hidden-charm tetraquark states are presented in Sec. \ref{secIII}.  In Sec. \ref{secIV}, the obtained results are summarized and discussed. For the sake of brevity, the expressions of the correlation function for the $Z_{c}(4020)^+$ tetraquark state and the distribution amplitudes of the photon are given in Appendix \ref{appenda} and \ref{appendb}, respectively.

\section{ QCD light-cone sum rules for the magnetic and quadrupole moments of the $Z_c$ states}\label{secII}

In this section, we explore the magnetic and quadrupole moment of hidden-charm tetraquark states composed of compact diquark-antidiquarks.  To do this, we write the two-point correlation function in the QCD sum rules in the presence of an external  electromagnetic background field as follows,
\begin{equation}
 \label{edmn01}
\Pi _{\mu \nu }(p,q)=i\int d^{4}xe^{ip\cdot x}\langle 0|\mathcal{T}\{J_{\mu}(x)
J_{\nu }^{\dagger }(0)\}|0\rangle_{\gamma}, 
\end{equation}
where  $\gamma$ represents the external electromagnetic background field, q is the momentum of the photon, and  $J_{\mu(\nu)}(x)$ is the interpolating current of the $Z_{c}$ states with quantum numbers $J^{P}=1^{+}$, which are given as 

\begin{align}
\label{curr}
    J_{\mu}^{Z_{c}}(x)&=\frac{\epsilon \tilde{\epsilon}}{\sqrt{2}}\Big\{
    [ { u^{bT}}(x) C \sigma_{\alpha\mu} \gamma_5 c^c(x)]
    [ \bar d^d(x)  \gamma^{\alpha} C { \bar c^{eT}}(x)]\nonumber\\ 
    &
    -[ { u^{bT}}(x) C \gamma^{\alpha}  c^c(x)]
    [ \bar d^d(x) \gamma_5  \sigma_{\alpha\mu} C  {\bar c^{eT}} (x)]
\Big\},
\end{align}
\begin{align}
\label{curr1}
   J_{\mu}^{Z^1_{c}}(x)&=\frac{\epsilon \tilde{\epsilon}}{\sqrt{2}}\Big\{
    [ { u^{bT}}(x) C \sigma_{\alpha\mu}  c^c(x)]
    [ \bar d^d(x) \gamma_5 \gamma^{\alpha} C { \bar c^{eT}}(x)] 
    \nonumber\\ 
    &
    +[ { u^{bT}}(x) C \gamma^{\alpha}\gamma_5  c^c(x)]
    [\bar d^d(x)   \sigma_{\alpha\mu} C  {\bar c^{eT}} (x)]
\Big\},
\end{align}
\begin{align}
\label{curr2}
    J_{\mu}^{Z^2_{c}}(x)&=\frac{\epsilon \tilde{\epsilon}}{\sqrt{2}}\Big\{
    [ { u^{bT}}(x) C \sigma_{\alpha\mu}  c^c(x)]
    [ \bar d^d(x) \gamma_5 \gamma^{\alpha} C { \bar c^{eT}}(x)] 
    \nonumber\\ 
    &
    -[ { u^{bT}}(x) C \gamma^{\alpha}\gamma_5  c^c(x)]
    [ \bar d^d(x)   \sigma_{\alpha\mu} C  {\bar c^{eT}} (x)]
\Big\},
  \end{align}
where  $\epsilon =\epsilon _{abc}$; $\tilde{\epsilon}=\epsilon _{ade}$; the $a$, $b$, $c$, $d$, and  $e$  are color indexes; $\sigma_{\mu\nu}=\frac{i}{2}[\gamma_{\mu},\gamma_{\nu}]$; and $C$ is the charge conjugation matrix.

Now, let us first calculate the hadronic side of the correlation function. 
The correlation function in Eq. ({\ref{edmn01}}) can be acquired by entering it into all intermediate hadronic sum rules, with the same quantum numbers as the corresponding interpolating currents $J_\mu$. After isolating the contributions of the ground $Z_c$ states, we obtain
 \begin{align}
\label{edmn04}
\Pi_{\mu\nu}^{Had} (p,q) &= \frac{1}{ [m_{Z_c}^2 - (p+q)^2][m_{Z_c}^2 - p^2]} \nonumber\\
\nonumber\\
& \times  \langle 0 \mid J_\mu (x) \mid Z_{c}(p, \varepsilon^\theta) \rangle \nonumber\\
\nonumber\\
&
\times 
\langle Z_{c}(p, \varepsilon^\theta) \mid Z_{c}(p+q, \varepsilon^\delta) \rangle_\gamma 
\nonumber\\
\nonumber\\
&
\times 
\langle Z_{c}(p+q,\varepsilon^\delta) \mid {J^\dagger}_\nu (0) \mid 0 \rangle
\nonumber\\
\nonumber\\
& +\cdot \cdot \cdot\,,
\end{align}
where dots denote the contributions coming the higher states and continuum. 
 
 The matrix elements of the interpolating current between one hadron and vacuum states in terms of the polarization vectors and residues are given as
\begin{align}
\langle 0 \mid J_\mu(x) \mid Z_{c}(p,\varepsilon^\theta) \rangle &= \lambda_{Z_{c}} \varepsilon_\mu^\theta\,,\\
\nonumber\\
\langle Z_{c}(p+q,\varepsilon^\delta) \mid {J^\dagger}_\nu (0) \mid 0 \rangle  &= \lambda_{Z_{c}} \varepsilon_\nu^{*\delta}\,.
\end{align}

The radiative transition matrix element in  Eq. (\ref{edmn04}) is written in terms of three Lorentz invariant form factors $G_1(Q^2)$,  $G_2(Q^2)$,  and $G_3(Q^2)$ as
\begin{align}
\langle Z_{c}(p,\varepsilon^\theta) \mid  Z_{c} (p+q,\varepsilon^{\delta})\rangle_\gamma &= - \varepsilon^\tau (\varepsilon^{\theta})^\alpha (\varepsilon^{\delta})^\beta \Big\{ 
\nonumber\\
\nonumber\\
& \times  G_1(Q^2) (2p+q)_\tau ~g_{\alpha\beta} 
\nonumber\\
\nonumber\\
& 
+ G_2(Q^2) ( g_{\tau\beta}~ q_\alpha -  g_{\tau\alpha}~ q_\beta) 
\nonumber\\ 
\nonumber\\
&- \frac{1}{2 m_{Z_{c}}^2} G_3(Q^2)~ (2p+q)_\tau 
\nonumber\\
\nonumber\\
& \times 
q_\alpha q_\beta  \Big\},\label{edmn06}
\end{align}
where $\varepsilon^\tau$  and $\varepsilon^{\delta(\theta)}$ are the polarization vectors of the photon and $Z_c$ states, respectively.

Employing Eqs.~(\ref{edmn04})-(\ref{edmn06}), the hadronic part of the correlation function becomes
%

\begin{align}
\label{edmn09}
 \Pi_{\mu\nu}^{Had}(p,q) &=  \frac{\varepsilon_\rho \, \lambda_{Z_c}^2}{ [m_{Z_c}^2 - (p+q)^2][m_{Z_c}^2 - p^2]}
 \bigg\{G_1(Q^2) \nonumber\\
& \times (2p+q)_\rho\bigg(g_{\mu\nu}-\frac{p_\mu p_\nu}{m_{Z_c}^2}
 -\frac{(p+q)_\mu (p+q)_\nu}{m_{Z_c}^2} \nonumber\\
 &+\frac{(p+q)_\mu p_\nu}{2m_{Z_c}^4} (Q^2+2m_{Z_c}^2)
 \bigg)\nonumber\\
&
 + G_2 (Q^2) \bigg(q_\mu g_{\rho\nu} - q_\nu g_{\rho\mu} -
\frac{p_\nu}{m_{Z_c}^2}  \big(q_\mu p_\rho  \nonumber\\
& - \frac{1}{2}
Q^2 g_{\mu\rho}\big) 
+
\frac{(p+q)_\mu}{m_{Z_c}^2}  \big(q_\nu (p+q)_\rho+ \frac{1}{2}
Q^2 g_{\nu\rho}\big)
\nonumber\\
&-  
\frac{(p+q)_\mu p_\nu p_\rho}{m_{Z_c}^4} \, Q^2
\bigg)\nonumber\\
&
-\frac{G_3(Q^2)}{m_{Z_c}^2}(2p+q)_\rho \bigg(
q_\mu q_\nu -\frac{p_\mu q_\nu}{2 m_{Z_c}^2} Q^2 \nonumber\\
&+\frac{(p+q)_\mu q_\nu}{2 m_{Z_c}^2} Q^2
-\frac{(p+q)_\mu q_\nu}{4 m_{Z_c}^4} Q^4\bigg)
\bigg\}\,.
\end{align}


The magnetic and quadrupole moments of hadrons are related to their magnetic and quadrupole form factors; more precisely, the magnetic and quadrupole moments are equal to the magnetic and quadrupole form factor at zero momentum square.  Magnetic ($F_M(Q^2)$) and quadrupole ($F_{\mathcal D} (Q^2)$) form factors, which are more directly accessible in experiments, are described via the form factors $G_1(Q^2)$,  $G_2(Q^2)$  and $G_3(Q^2)$ 
\begin{align}
\label{edmn07}
&F_M(Q^2) = G_2(Q^2)\,,\nonumber \\
&F_{\mathcal D}(Q^2) = G_1(Q^2)-G_2(Q^2)+(1+\lambda) G_3(Q^2)\,,
\end{align}
where  $\lambda=Q^2/4 m_{Z_c}^2$ with $Q^2=-q^2$.  At the static limit, that is, $Q^2 = 0 $, the form factors $F_M(Q^2=0)$ and $F_{\mathcal D}(Q^2=0)$ are proportional to the
 magnetic ($\mu_{Z_c}$) and quadrupole ($\mathcal {D}_{Z_c}$) moments in
the following way:
\begin{align}
\label{edmn08}
&e F_M(Q^2=0) = 2 m_{Z_c} \mu_{Z_c} \,, \nonumber\\
&e F_{\cal D}(Q^2=0) = m_{Z_c}^2 {\mathcal {D}_{Z_c}}\,.
\end{align}

Let us evaluate the QCD part of the correlation function. The QCD side of the above correlation function is computed considering the QCD degrees of freedom in the deep Euclidean region. To do this,  we must  insert the interpolating currents in Eqs. (\ref{curr})-(\ref{curr2}) into the correlation function. After substituting the explicit forms of the interpolating currents into the correlation function and applying contractions through Wick’s theorem, we obtain the QCD side as

\begin{widetext}
 
\begin{align}
\label{neweq}
\Pi _{\mu \nu }^{\mathrm{QCD-Z_{c}}}(p,q)&=\frac{ \epsilon \tilde{\epsilon} \epsilon^{\prime} \tilde{\epsilon}^{\prime}}{2}
\int d^{4}xe^{ipx} \langle 0 | \Big\{ 
\mathrm{Tr}\Big[\gamma^{\alpha}{\tilde S}_{c}^{e^{\prime }e}(-x)\gamma ^{\beta}S_{d}^{d^{\prime }d}(-x)\Big] 
\mathrm{Tr}\Big[\sigma_{\mu\alpha}\gamma _{5 }{S}_{c}^{cc^{\prime }}(x)\gamma _{5}\sigma_{\nu\beta}\tilde S_{u}^{bb^{\prime }}(x)\Big] \notag \\
&-\mathrm{Tr}\Big[ \gamma^{\alpha}{\tilde S}_{c}^{e^{\prime }e}(-x)\gamma _{5}\sigma_{\nu\beta}S_{d}^{d^{\prime }d}(-x)\Big]   
\mathrm{Tr}\Big[ \sigma_{\mu\alpha}\gamma_{5 }{S}_{c}^{cc^{\prime }}(x)\gamma^{\beta}\tilde S_{u}^{bb^{\prime }}(x)\Big] \notag \\
&-\mathrm{Tr}\Big[\sigma_{\mu\alpha}\gamma _{5}{\tilde S}_{c}^{e^{\prime }e}(-x)\gamma^{\beta }S_{d}^{d^{\prime }d}(-x)\Big]    
\mathrm{Tr}\Big[ \gamma^{\alpha}{S}_{c}^{cc^{\prime }}(x)\gamma_{5}\sigma_{\nu\beta}\tilde S_{u}^{bb^{\prime }}(x)\Big] \notag \\
&+\mathrm{Tr}\Big[\sigma_{\mu\alpha}\gamma_{5 }{\tilde S}_{c}^{e^{\prime }e}(-x)\gamma _{5}\sigma_{\nu\beta}S_{d}^{d^{\prime }d}(-x)\Big]  
\mathrm{Tr}\Big[\gamma^{\alpha}{S}_{c}^{cc^{\prime }}(x) \gamma^{\beta}\tilde S_{u}^{bb^{\prime }}(x)\Big]
 \Big\}| 0 \rangle_\gamma,
\end{align} 
\begin{align}
\label{neweq1}
\Pi _{\mu \nu }^{\mathrm{QCD-Z_{c}^1}}(p,q)&=\frac{ \epsilon \tilde{\epsilon} \epsilon^{\prime} \tilde{\epsilon}^{\prime}}{2}
\int d^{4}xe^{ipx} \langle 0 | \Big\{ 
\mathrm{Tr}\Big[ \gamma _{5 }\gamma^{\alpha}{\tilde S}_{c}^{e^{\prime }e}(-x)\gamma ^{\beta} \gamma _{5 }S_{d}^{d^{\prime }d}(-x)\Big] 
\mathrm{Tr}\Big[\sigma_{\mu\alpha}{S}_{c}^{cc^{\prime }}(x)\sigma_{\nu\beta}\tilde S_{u}^{bb^{\prime }}(x)\Big] \notag \\
&+\mathrm{Tr}\Big[ \gamma _{5 } \gamma^{\alpha}{\tilde S}_{c}^{e^{\prime }e}(-x)\sigma_{\nu\beta}S_{d}^{d^{\prime }d}(-x)\Big]   
\mathrm{Tr}\Big[ \sigma_{\mu\alpha}{S}_{c}^{cc^{\prime }}(x)\gamma^{\beta} \gamma_{5 }\tilde S_{u}^{bb^{\prime }}(x)\Big] \notag \\
&+\mathrm{Tr}\Big[\sigma_{\mu\alpha}{\tilde S}_{c}^{e^{\prime }e}(-x)\gamma^{\beta } \gamma _{5}S_{d}^{d^{\prime }d}(-x)\Big]    
\mathrm{Tr}\Big[ \gamma _{5}\gamma^{\alpha}{S}_{c}^{cc^{\prime }}(x)\sigma_{\nu\beta}\tilde S_{u}^{bb^{\prime }}(x)\Big] \notag \\
&+\mathrm{Tr}\Big[\sigma_{\mu\alpha}{\tilde S}_{c}^{e^{\prime }e}(-x)\sigma_{\nu\beta}S_{d}^{d^{\prime }d}(-x)\Big]  
\mathrm{Tr}\Big[\gamma_{5} \gamma^{\alpha}{S}_{c}^{cc^{\prime }}(x) \gamma^{\beta} \gamma_{5}\tilde S_{u}^{bb^{\prime }}(x)\Big]
 \Big\}| 0 \rangle_\gamma,
\end{align} 
\begin{align}
\label{neweq2}
\Pi _{\mu \nu }^{\mathrm{QCD-Z_{c}^2}}(p,q)&=\frac{ \epsilon \tilde{\epsilon} \epsilon^{\prime} \tilde{\epsilon}^{\prime}}{2}
\int d^{4}xe^{ipx} \langle 0 | \Big\{ 
\mathrm{Tr}\Big[ \gamma _{5 }\gamma^{\alpha}{\tilde S}_{c}^{e^{\prime }e}(-x)\gamma ^{\beta} \gamma _{5 }S_{d}^{d^{\prime }d}(-x)\Big] 
\mathrm{Tr}\Big[\sigma_{\mu\alpha}{S}_{c}^{cc^{\prime }}(x)\sigma_{\nu\beta}\tilde S_{u}^{bb^{\prime }}(x)\Big] \notag \\
&-\mathrm{Tr}\Big[ \gamma _{5 } \gamma^{\alpha}{\tilde S}_{c}^{e^{\prime }e}(-x)\sigma_{\nu\beta}S_{d}^{d^{\prime }d}(-x)\Big]   
\mathrm{Tr}\Big[ \sigma_{\mu\alpha}{S}_{c}^{cc^{\prime }}(x)\gamma^{\beta} \gamma_{5 }\tilde S_{u}^{bb^{\prime }}(x)\Big] \notag \\
&-\mathrm{Tr}\Big[\sigma_{\mu\alpha}{\tilde S}_{c}^{e^{\prime }e}(-x)\gamma^{\beta } \gamma _{5}S_{d}^{d^{\prime }d}(-x)\Big]    
\mathrm{Tr}\Big[ \gamma _{5}\gamma^{\alpha}{S}_{c}^{cc^{\prime }}(x)\sigma_{\nu\beta}\tilde S_{u}^{bb^{\prime }}(x)\Big] \notag \\
&+\mathrm{Tr}\Big[\sigma_{\mu\alpha}{\tilde S}_{c}^{e^{\prime }e}(-x)\sigma_{\nu\beta}S_{d}^{d^{\prime }d}(-x)\Big]  
\mathrm{Tr}\Big[\gamma_{5} \gamma^{\alpha}{S}_{c}^{cc^{\prime }}(x) \gamma^{\beta} \gamma_{5}\tilde S_{u}^{bb^{\prime }}(x)\Big]
 \Big\}| 0 \rangle_\gamma,
\end{align} 

\end{widetext}
where $S_{c}(x)$ and $S_{q}(x)$ represent propagators of  heavy and light quarks. The explicit forms of the quark propagators are written as~\cite{Yang:1993bp, Belyaev:1985wza}

\begin{align}
\label{edmn12}
S_{q}(x) &= S_q^{free}
- \frac{\langle \bar qq \rangle }{12} \Big(1-i\frac{m_{q} \xslash}{4}   \Big)
- \frac{\langle \bar q \sigma.G q \rangle }{192}x^2  \Big(1 \nonumber\\
&-i\frac{m_{q} \xslash}{6}   \Big)
-\frac {i g_s }{32 \pi^2 x^2} ~G^{\mu \nu} (x) \bigg[\rlap/{x}
\sigma_{\mu \nu} +  \sigma_{\mu \nu} \rlap/{x}
 \bigg],
\end{align}%
and
\begin{align}
\label{edmn13}
S_{c}(x)&=S_c^{free}
-\frac{g_{s}m_{c}}{16\pi ^{2}} \int_0^1 dv\, G^{\mu \nu }(vx)\Bigg[ (\sigma _{\mu \nu }{\xslash}
  +{\xslash}\sigma _{\mu \nu })
  \nonumber\\
& \times 
  \frac{K_{1}\Big( m_{c}\sqrt{-x^{2}}\Big) }{\sqrt{-x^{2}}}
+2\sigma_{\mu \nu }K_{0}\Big( m_{c}\sqrt{-x^{2}}\Big)\Bigg],
\end{align}%
where 
\begin{align}
S_q^{free} &=\frac{1}{2 \pi^2 x^2}\Big( i \frac{{\xslash}}{x^{2}}-\frac{m_{q}}{2 } \Big),\\
\nonumber\\
S_c^{free} &= \frac{m_{c}^{2}}{4 \pi^{2}} \Bigg[ \frac{K_{1}\Big(m_{c}\sqrt{-x^{2}}\Big) }{\sqrt{-x^{2}}}
+i\frac{{\xslash}~K_{2}\Big( m_{c}\sqrt{-x^{2}}\Big)}
{(\sqrt{-x^{2}})^{2}}\Bigg].
\end{align} 

The correlation functions in Eqs.~(\ref{neweq})-(\ref{neweq2}) contain short distance (perturbative), and long distance (nonperturbative) contributions. To obtain the expressions of the contributions when the photon is radiated at a short distance, it is adequate to modify one of the propagators in Eqs.~(\ref{neweq})-(\ref{neweq2}) as follows

\begin{align}
\label{free}
S^{free}(x) \rightarrow \int d^4y\, S^{free} (x-y)\,\rlap/{\!A}(y)\, S^{free} (y)\,,
\end{align}
where the remaining three propagators in Eqs. (\ref{neweq})-(\ref{neweq2}) are considered free propagators. This amounts to taking $\bar T_4^{\gamma} (\underline{\alpha}) = 0$ and $S_{\gamma} (\underline {\alpha}) = \delta(\alpha_{\bar q})\delta(\alpha_{q})$ as the light-cone distribution amplitude in the three particle distribution amplitudes (see Ref. \cite{Li:2020rcg}). 

To obtain the expressions for when the photon is radiated at a long distance the correlation function can be acquired from Eqs. (\ref{neweq})-(\ref{neweq2}) by substituting one of the u/d-quark propagators via

\begin{align}
\label{edmn14}
S_{\mu\nu}^{ab}(x) \rightarrow -\frac{1}{4} \big[\bar{q}^a(x) \Gamma_i q^b(0)\big]\big(\Gamma_i\big)_{\mu\nu},
\end{align}
 where   $\Gamma_i = I, \gamma_5, \gamma_\mu, i\gamma_5 \gamma_\mu, \sigma_{\mu\nu}/2$. In this approach, the three remaining quark propagators are considered full quark propagators containing both perturbative and non-perturbative contributions.  When a photon interacts with light-quark fields nonperturbatively the matrix elements of the nonlocal operators $\langle \gamma(q)\vel \bar{q}(x) \Gamma_i G_{\mu\nu}q(0) \ver 0\rangle$  and $\langle \gamma(q)\vel \bar{q}(x) \Gamma_i q(0) \ver 0\rangle$appear between the photon state and vacuum, which are parameterized in terms of photon distribution amplitudes (DAs) (for details see Ref. \cite{Ball:2002ps}).  Together with these matrix elements non-local operators such as four quarks ($\bar qq \bar q q$) and two gluons ($\bar q G G q$)  are expected to appear. However, it is known that the contributions of such terms are small, which is confirmed by conformal spin expansion \cite{Balitsky:1987bk,Braun:1989iv}, and hence we neglect them.   The QCD side of the correlation function is evaluated by employing Eqs.~(\ref {neweq})-(\ref {edmn14}).  The Fourier transform is then used to transfer the x-space expressions to the momentum space.

QCD sum rules for the hadron parameters are obtained by equating the correlation functions evaluated at both the hadronic and quark-gluon parameters through quark-hadron duality. Then we choose the structure $(\varepsilon.p) (p_\mu q_\nu -p_\nu q_\mu) $ and $(\varepsilon.p) q_\mu q_\nu$ for the magnetic and quadrupole moments, respectively. As a result, we obtain the following: 

\begin{widetext}
 
\begin{align}
 \mu_{Z_{c}} &=  \frac{e^{\frac{m_{Z_{c}}^2}{M^2}}}{\lambda_{Z_{c}}^2}  \,\, \Delta_1^{QCD}(M^2,s_0), ~~~~~~\mathcal{D}_{Z_c}  = \frac{m_{Z_{c}}^2 e^{\frac{m_{Z_{c}}^2}{M^2}}}{\lambda_{Z_{c}}^2}  \,\, \Delta_2^{QCD}(M^2,s_0), \label{sonj1} \\
 \mu_{Z^1_{c}} &= \frac{e^{\frac{m_{Z^1_{c}}^2}{M^2}}}{\lambda_{Z^1_{c}}^2}  \,\, \Delta_3^{QCD}(M^2,s_0), ~~~~~~\mathcal{D}_{Z_c^1}  = \frac{m_{Z_{c}^1}^2 e^{\frac{m_{Z_{c}^1}^2}{M^2}}}{\lambda_{Z_{c}^1}^2} \,\, \Delta_4^{QCD}(M^2,s_0),\\
  \mu_{Z^2_{c}} &= \frac{e^{\frac{m_{Z^2_{c}}^2}{M^2}}}{\lambda_{Z^2_{c}}^2} \,\, \Delta_5^{QCD}(M^2,s_0), \,\, ~~~~~~\mathcal{D}_{Z_c^2}  = \frac{m_{Z_{c}^2}^2 e^{\frac{m_{Z_{c}^2}^2}{M^2}}}{\lambda_{Z_{c}^2}^2}  \,\, \Delta_6^{QCD}(M^2,s_0),\label{sonj2}
 \end{align}
 
 \end{widetext}
where $M^2$ is the Borel mass and $s_0$ is the continuum threshold parameter. For the sake of simplicity, only the explicit expressions of the $\Delta_1^{QCD}(M^2,s_0)$ function are presented in Appendix \ref{appenda}, because the remaining functions are in a similar form.

\section{Numerical analysis} \label{secIII}
 
The section contains a numerical analysis
for the magnetic and quadrupole moments of the $Z_c$ states. 
The following QCD parameters are used in our calculations:   $m_u=m_d=0$, 
$m_c = (1.27\pm 0.02)\,$GeV, 
$m_{Z_c}= 4024.1 \pm 1.9$~MeV \cite{ParticleDataGroup:2022pth}, $m_{Z_c^1}= 4051 \pm 14 ^{+20}_{-41}$~MeV \cite{ParticleDataGroup:2022pth}, $m_{Z_c^2}= 4600$~MeV \cite{LHCb:2019maw},
$\langle \bar uu\rangle $ = $\langle \bar dd\rangle$=$(-0.24\pm0.01)^3\,$GeV$^3$  \cite{Ioffe:2005ym},   
$\langle g_s^2G^2\rangle = 0.88~ $GeV$^4$~\cite{Matheus:2006xi}, and $f_{3\gamma}=-0.0039~$GeV$^2$~\cite{Ball:2002ps}. From Eqs. (\ref{sonj1})-(\ref{sonj2}), it follows that for the determination of magnetic and quadrupole moments, the residues of the $Z_c$ states are needed. These residues are calculated in  Ref.~\cite{Wang:2019tlw}, and the values are obtained as $\lambda_{Z_c}= (6.09 \pm 0.90)\times 10^{-2}$ GeV$^5$, $\lambda_{Z_c^1}= (3.67 \pm 0.67)\times 10^{-2}$ GeV$^5$ and $\lambda_{Z_c^2}= (1.18 \pm 0.21)\times 10^{-1}$ GeV$^5$.  The photon DAs are one of the main non-perturbative inputs of the QCD light-cone sum rules. A comprehensive analysis of the photon DAs was conducted in Ref.~\cite{Ball:2002ps}. For completeness, the parameters used in the photon DAs are presented in Appendix \ref{appendb}.

\begin{widetext}

\begin{figure}[t]
\centering
 \includegraphics[width=0.45\textwidth]{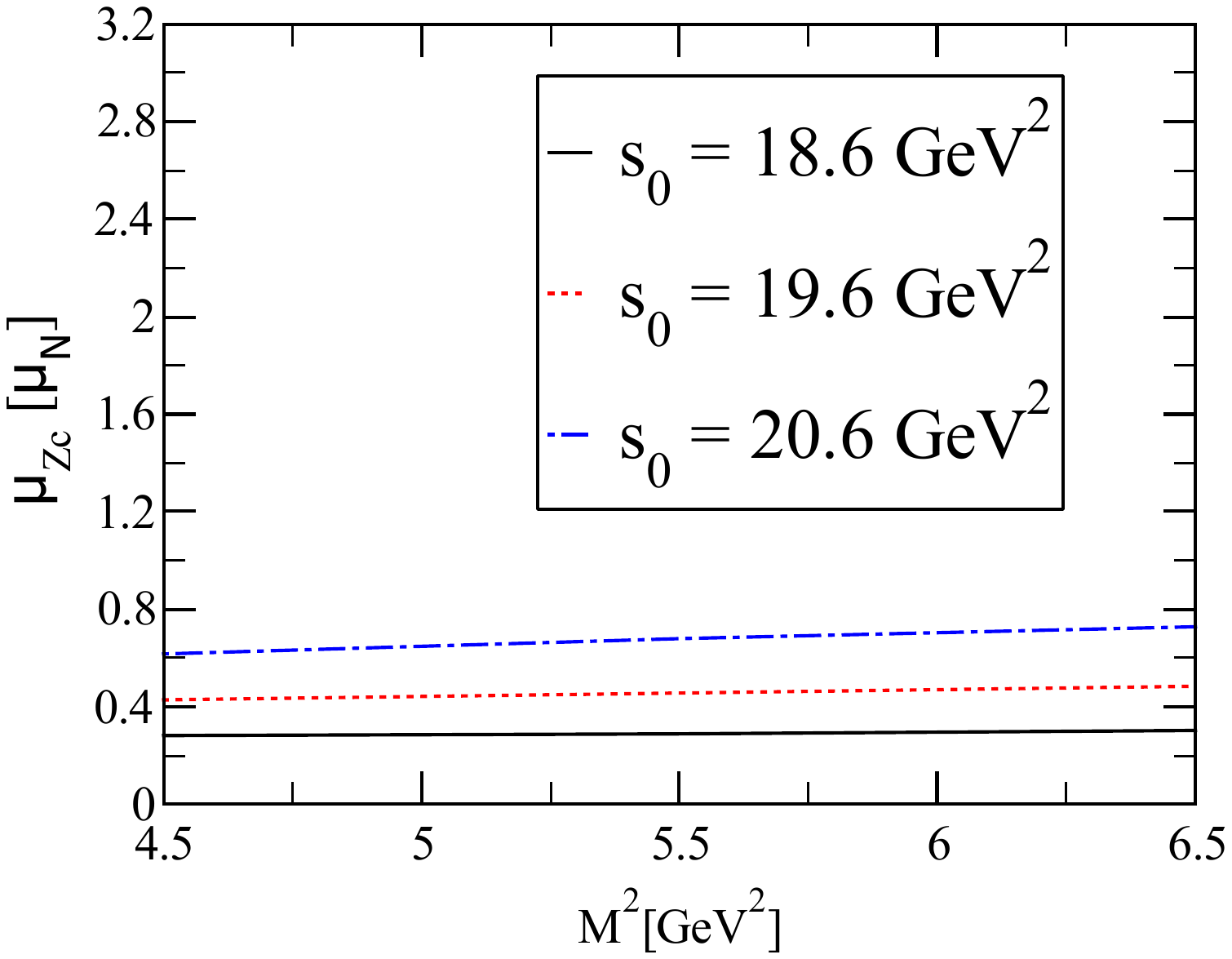}~~~~
 \includegraphics[width=0.45\textwidth]{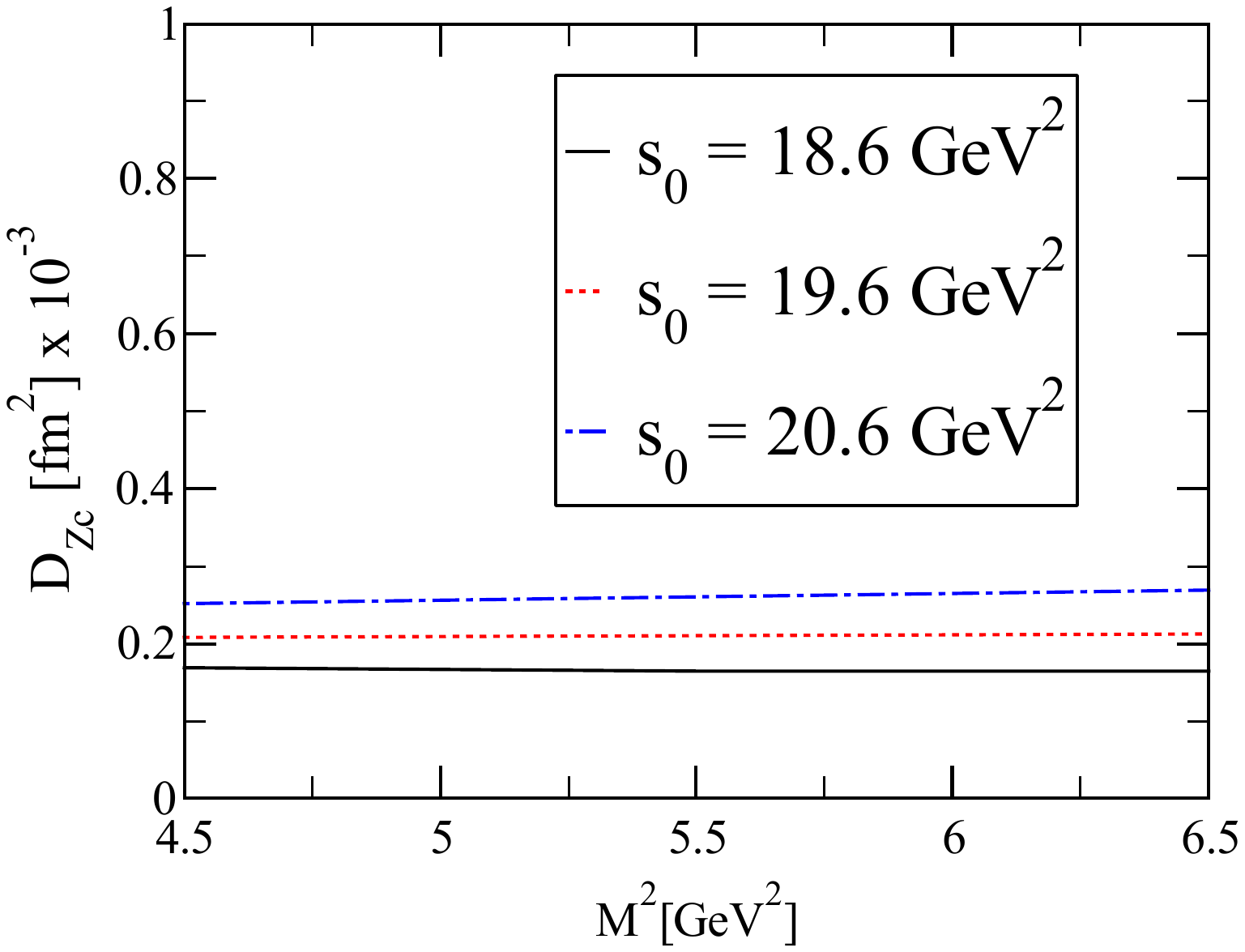}\\
 \includegraphics[width=0.45\textwidth]{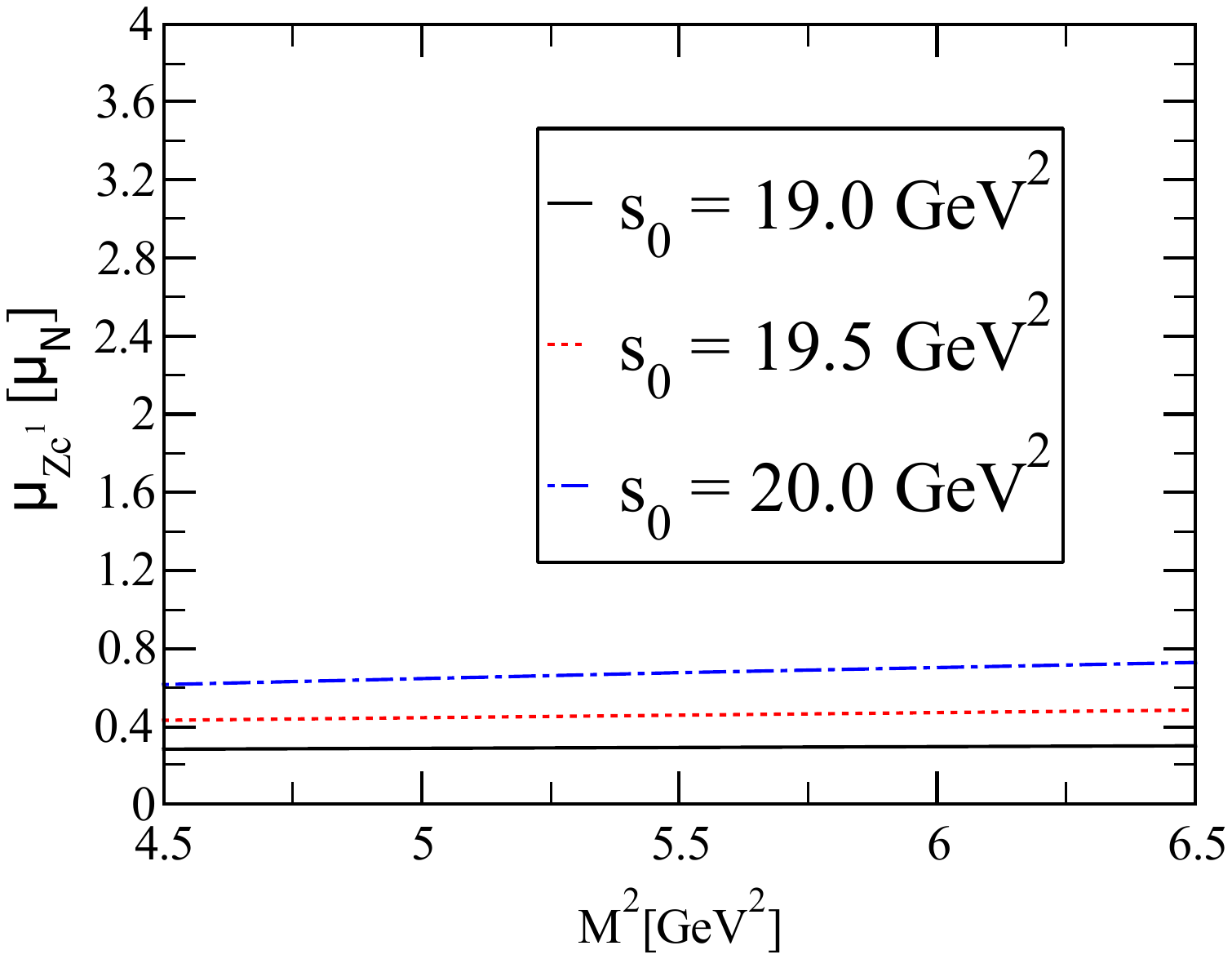}~~~~
 \includegraphics[width=0.45\textwidth]{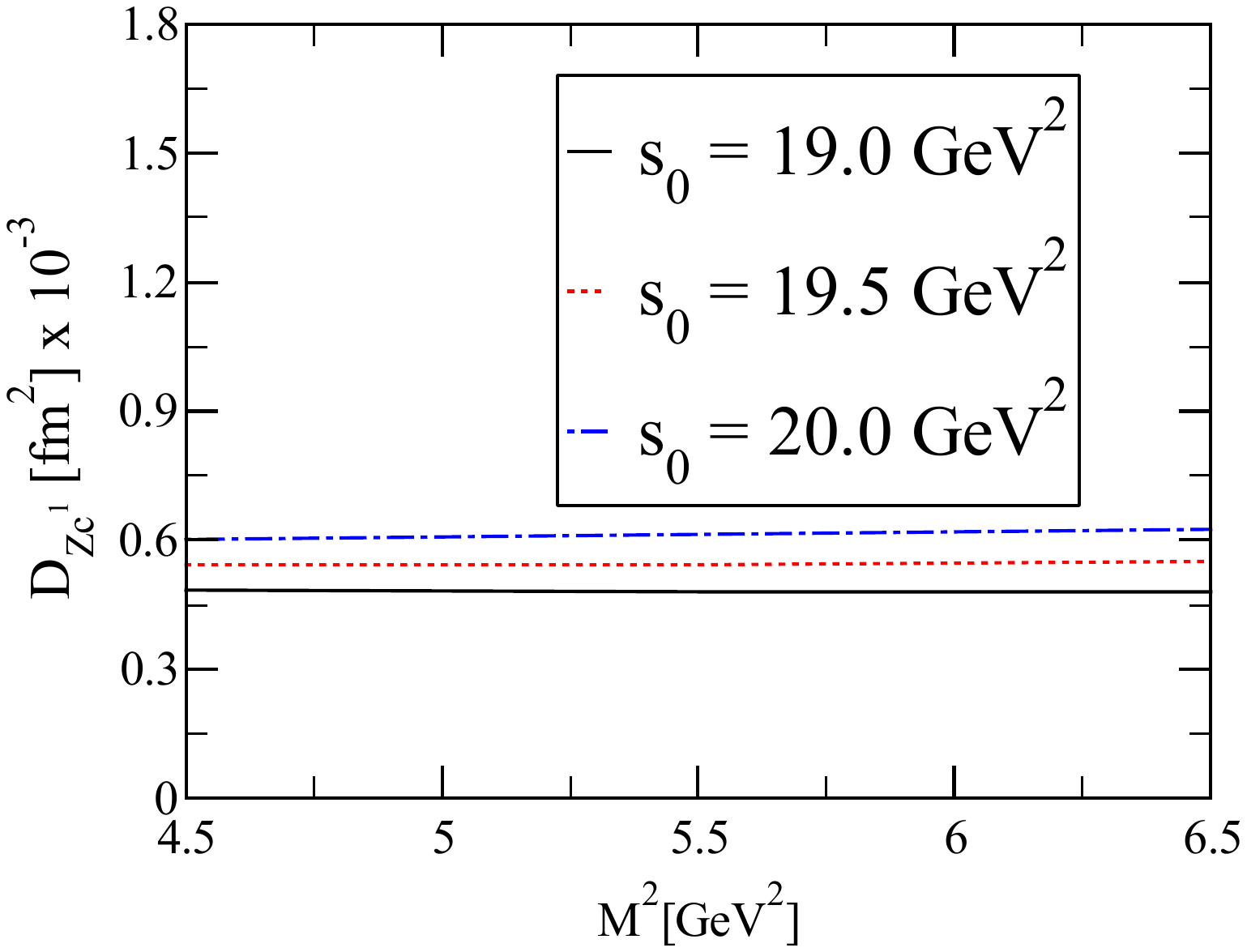}\\
 \includegraphics[width=0.45\textwidth]{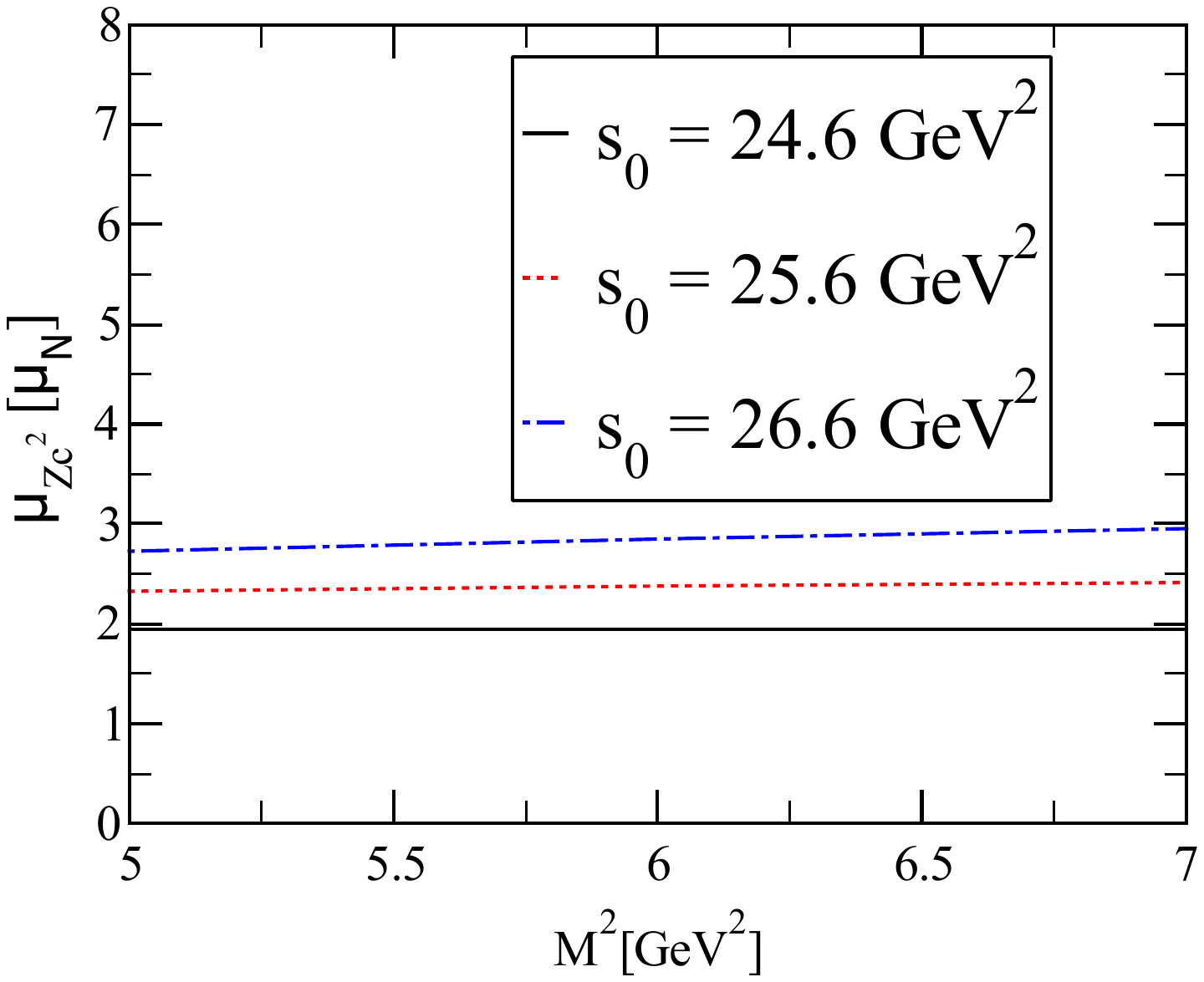}~~~~
 \includegraphics[width=0.45\textwidth]{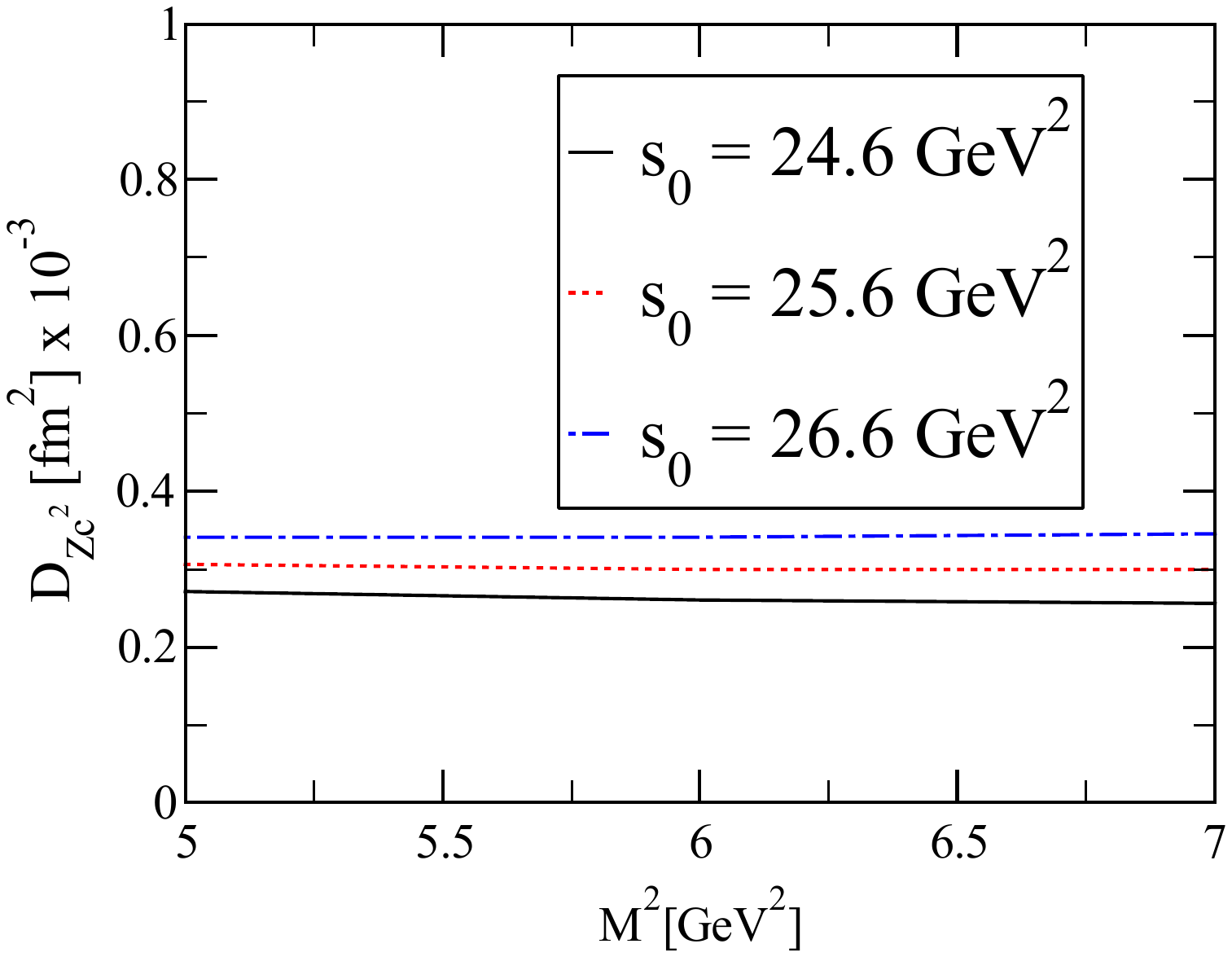}\\
 \caption{ Dependencies of the magnetic and quadrupole moments of the hidden-charm tetraquark states on $M^{2}$ at three different values of $s_0$.}
 \label{Msqfig}
  \end{figure}
  
  \end{widetext}

The sum rules also depend on the helping parameters, as mentioned in the previous section, that is,  Borel mass square parameter $M^2$ and continuum threshold $s_0$. The physical observables, namely the magnetic and quadrupole moments, should be independent of these parameters. Therefore, we search for the  working regions of these additional parameters in such a way that, in these regions, the magnetic and quadrupole moments are nearly independent of these parameters. In determining the working regions of the parameters $M^2$ and $s_0$, the standard prescription of the technique used, the operator product expansion (OPE) convergence, and the pole contribution (PC) dominance are considered. To characterize the above constraints, it is convenient to use the following equations:
\begin{align}
 \mbox{PC} =\frac{\Delta (M^2,s_0)}{\Delta (M^2,\infty)} \geq  30\%,
 \end{align}
and 
\begin{align}
 \mbox{OPE Convergence} =\frac{\Delta^{\mbox{Dim7}} (M^2,s_0)}{\Delta (M^2,s_0)}\leq  5\%,
 \end{align}
 where $\Delta^{\mbox{Dim7}} (M^2,s_0)$ represents the contribution of the highest dimensional term in  OPE.  The OPE convergence and PC values for each state, along with working regions acquired for $M^2$ and $s_0$ are presented in Table \ref{parameter}. From the values given in Table \ref{parameter}, we find that the working regions determined for $M^2$ and $s_0$ meet the above requirements. Having determined the working regions of $M^2$ and $s_0$, we now study the dependence of magnetic and quadrupole moments on $M^2$ at several fixed values of $s_0$.  From Fig. \ref{Msqfig}, we notice that indeed magnetic and quadrupole moments represent good stability regarding the variation in  $M^2$ in its working region.

\begin{widetext}

\begin{table}[htp]
	\addtolength{\tabcolsep}{10pt}
	\caption{Working windows of  $s_0$ and  $M^2$ along with the PC and OPE convergence for the magnetic and quadrupole moments of hidden-charm tetraquark states.}
	\label{parameter}
		\begin{center}
\begin{tabular}{l|ccccc}
                \hline\hline
                \\
~~~~~State~~ & $s_0$ (GeV$^2$)& 
$M^2$ (GeV$^2$) & ~~  PC ($\%$) ~~ &  OPE Convergence 
 ($\%$) \\
\\
                                        \hline\hline
                                        \\
   ~~ $Z_c(4020)^+$ ~~                     & $18.6-20.6$ & $4.5-6.5$ & $34-60$ &  $2.8$  
                        \\
                        \\
  ~~  $Z_c(4050)^+$ ~~                   & $19.0-20.0$ & $4.5-6.5$ & $31-57$ &  $3.2$  \\
                        \\
   ~~   $Z_c(4600)^+$  ~~                &  $24.6-26.6$ & $5.0-7.0$ & $33-60$ & $ 3.0$   \\
   \\
                                       \hline\hline
 \end{tabular}
\end{center}
\end{table}
\end{widetext}

Our final results for the magnetic and quadrupole moments of the $Z_{c}(4020)^+$, $Z_{c}(4050)^+$, and $Z_{c}(4600)^{+}$ states are presented in Table \ref{sonuc}. The uncertainties result from the variation in the Borel parameter $M^{2}$ and the continuum threshold $s_0$ as well as from uncertainties in the input parameters.  The magnitude of the numerical results of the magnetic moments also allows them to be measured experimentally. The magnetic moments of these states are sufficiently large to be measured in future experiments by examining the magnetic moment results. We obtain a non-zero, but small, value for the quadrupole moments of the $Z_c$ states, which represents a non-spherical charge distribution. The sign of the quadrupole moments is positive for the $Z_c$ tetraquark states, which correspond to  prolate charge distributions.
\begin{table}[htp]
	\addtolength{\tabcolsep}{10pt}
	\caption{ Results of the magnetic and quadrupole moments of hidden-charm tetraquark states.}
	\label{sonuc}
		\begin{center}
\begin{tabular}{l|ccccc}
                \hline\hline
                \\
~~~~~State~~ & $\mu_{Z_c}$[$\mu_N$]&  $\mathcal{D}_{Z_c}[\mbox{fm}^2] \times 10^{-3} $  \\
\\
                                        \hline\hline
                                        \\
   ~~ $Z_c(4020)^+$ ~~& $0.50 ^{+0.22}_{-0.22}$ & $0.20 ^{+0.05}_{-0.04}$   
                        \\
                        \\
  ~~  $Z_c(4050)^+$ ~~& $1.22 ^{+0.34}_{-0.32}$ & $0.57 ^{+0.07}_{-0.08}$   \\
                        \\
   ~~ $Z_c(4600)^+$ ~~&  $2.40 ^{+0.53}_{-0.48}$ & $0.30 ^{+0.05}_{-0.04}$    \\
   \\

                                       \hline\hline
 \end{tabular}
\end{center}
\end{table}

Before concluding this section, we must remark on the measurement of the magnetic moments of unstable hadrons. The short lifetimes of the $Z_{c}$ states make it difficult to measure the magnetic moment at current experimental facilities. However, the accumulation of more data from different experiments may make this possible in the future.  The $\Delta^+(1232)$ baryon also has a very short lifetime: however,  its magnetic moment is obtained from experimental data on the $\gamma N \rightarrow \Delta \rightarrow \Delta \gamma \rightarrow \pi N \gamma $ process \cite{Pascalutsa:2004je, Pascalutsa:2005vq, Pascalutsa:2007wb}.
Therefore, a technique for the determination of the magnetic and higher multipole moments is based on the soft photon emission of hadrons. This technique is recommended in Ref.~\cite{Zakharov:1968fb}. The photon also carries information about the magnetic and higher multipole moments of the emitted hadron. The radiative transition matrix element with respect to the energy of the photon can be described as 
\begin{align}
 M \sim A\,(E_\gamma)^{-1} + B\,(E_\gamma)^0 + C\, E_\gamma +\cdots \,,
\end{align}
where $E_\gamma$ is the energy of the photon. 
The electric charge contributes to the amplitude in the order of $(E_\gamma )^{-1}$, whereas the magnetic moment contribution is defined by the term $(E_\gamma)^0$. Thus, the magnetic moment of a hadron can be determined by measuring the decay width or cross-section of the radiative transition process and ignoring the small contributions of the linear and higher order terms in $E_\gamma$.

\section{Discussion and concluding remarks}\label{secIV}

The magnetic and quadrupole moments of the $Z_{c}(4020)^+$, $Z_{c}(4050)^+$ and $Z_{c}(4600)^{+}$ states, assuming that these states are represented as compact diquark-antidiquark states with $J^P = 1^+$ quantum numbers, are calculated in the framework of the QCD light-cone sum rules method.  Magnetic and quadrupole moments represent one of the most promising classes of decay for obtaining data on the electromagnetic properties that are prominent in revealing the internal structure of hadrons. Measuring the magnetic and quadrupole moments of hidden-charm tetraquark states in future experimental facilities will be helpful in determining the quantum numbers and in understanding the substructure of these hidden-charm tetraquark states. In addition, we hope that the magnetic and quadrupole moments of the $Z_{c}(4020)^+$, $Z_{c}(4050)^+$, and $Z_{c}(4600)^{+}$ states can be calculated using other approaches in the future, and these investigations will contribute to our knowledge of the magnetic and quadrupole moments of the $Z_{c}(4020)^+$, $Z_{c}(4050)^+$, and $Z_{c}(4600)^{+}$ states.

\begin{widetext}

\appendix
\section{Explicit forms of the $\Delta_1^{QCD}(M^2,s_0)$ function } \label{appenda} 
In this appendix, we present the explicit expressions of the analytical expressions obtained for the magnetic moment of the $Z_c(4020)^+$ state as follows: 
\begin{align}
\Delta_1^{QCD}(M^2,s_0)&=  \frac {27 (e_d - e_u+e_c)} {655360 \pi^5}\Bigg[
   I[0, 5, 3, 1] - 3 I[0, 5, 3, 2] + 3 I[0, 5, 3, 3] - 
    I[0, 5, 3, 4] - 3 I[0, 5, 4, 1] + 6 I[0, 5, 4, 2] \nonumber\\
    &- 
    3 I[0, 5, 4, 3] + 3 I[0, 5, 5, 1] - 3 I[0, 5, 5, 2] - 
    I[0, 5, 6, 1]\Bigg]\nonumber\\
    &+\frac {m_c^2 (e_d - e_u)  } {32768 \pi^5}\Big (I[0, 4, 2, 2] - 
   2 I[0, 4, 2, 3] + I[0, 4, 2, 4] - 2 I[0, 4, 3, 2] + 
   2 I[0, 4, 3, 3] + I[0, 4, 4, 2]\Big)
   \nonumber\\
    &+\frac {m_c P_ 1 P_ 2 } {7077888 \pi^3} \Bigg[\Big (234 e_u I_3[\mathcal S] + 120 e_u I_ 3[\mathcal {\tilde S}] + 
       195 e_d I_ 4[\mathcal S] + 82 e_d I_ 4[\mathcal {\tilde S}] + 
       384 (e_d - e_u) I_ 6[h_{\gamma}]\Big)
       \nonumber I[0, 1, 3, 0]\Bigg]\nonumber\\
        &-\frac {P_ 1 f_{3\gamma} m_c^2 } {110592 \pi^3} (e_d - e_u)   \Big(48 I[0, 1, 2, 0] + 
   I[0, 1, 3, 0]\Big) I_6[\psi^{\nu}]\nonumber\\
        &
   +\frac {P_1 f_ {3 \gamma }} {9437184 \pi^3} \Big (e_u I_1[\mathcal A] + 6 e_u I_ 1[\mathcal V] + 
     e_d (I_ 2[\mathcal A] + 6 I_ 2[\mathcal V]) + 
     256 (-e_d + e_u) I_ 6[\psi^{\nu}]\Big) I[0, 2, 4, 0]\nonumber
    \end{align}
   \begin{align}
    &
    +\frac {m_c P_ 2} {393216 \pi^3} \Bigg[-4 \Big(10 e_u I_ 3[\mathcal S] + 
       38 e_u I_ 3[\mathcal {\tilde S}] + 6 e_d I_ 4[\mathcal S] + 
       45 e_d I_ 4[\mathcal {\tilde S}]\Big) I[0, 3, 4, 0]      + 
    3 \Big (4 e_d I_ 4[\mathcal S] + 
       5 e_d I_ 4[\mathcal {\tilde S}] \nonumber\\
       &+ 
       6 e_u (I_ 3[\mathcal S] + I_ 3[\mathcal {\tilde S}] - 
          32 I_ 6[h_ {\gamma}])  + 
       96 (2e_d - e_u )I_ 6[h_ {\gamma}]\Big) I[0, 3, 5, 
       0]\Bigg]\nonumber\\
       &
       +\frac {f_ {3 \gamma}} {2097152 \pi^3}\Bigg[
   4  \Big (22 e_u I_ 1[\mathcal A] - 25 e_u I_ 1[\mathcal V] + 
       22 e_d I_ 2[\mathcal A] - 25 e_d I_ 2[\mathcal V]\Big) I[0, 4, 
      5, 0] - 3 \Big (6 e_u I_ 1[\mathcal A]  + e_u I_ 1[\mathcal V] \nonumber\\
    &+ 
       e_d (6 I_ 2[\mathcal A] + I_ 2[\mathcal V]) + 
       448 (e_d - e_u) I_ 6[\psi^{\nu}]\Big) I[0, 4, 6, 
       0]\Bigg],
        \end{align}
where $P_1 =\langle g_s^2 G^2\rangle$ and  $P_2 =\langle \bar q q \rangle$  are gluon and u/d-quark condensates, respectively.  
The functions~$I[n,m,l,k]$, $I_1[\mathcal{A}]$,~$I_2[\mathcal{A}]$,~$I_3[\mathcal{A}]$,~$I_4[\mathcal{A}]$,
~$I_5[\mathcal{A}]$, and ~$I_6[\mathcal{A}]$ are
defined as
\begin{align}
 I[n,m,l,k]&= \int_{4 m_c^2}^{s_0} ds \int_{0}^1 dt \int_{0}^1 dw~ e^{-s/M^2}~
 s^n\,(s-4\,m_c^2)^m\,t^l\,w^k,\nonumber\\
  I_1[\mathcal{A}]&=\int D_{\alpha_i} \int_0^1 dv~ \mathcal{A}(\alpha_{\bar q},\alpha_q,\alpha_g)
 \delta'(\alpha_ q +\bar v \alpha_g-u_0),\nonumber\\
  I_2[\mathcal{A}]&=\int D_{\alpha_i} \int_0^1 dv~ \mathcal{A}(\alpha_{\bar q},\alpha_q,\alpha_g)
 \delta'(\alpha_{\bar q}+ v \alpha_g-u_0),\nonumber\\
   I_3[\mathcal{A}]&=\int D_{\alpha_i} \int_0^1 dv~ \mathcal{A}(\alpha_{\bar q},\alpha_q,\alpha_g)
 \delta(\alpha_ q +\bar v \alpha_g-u_0),\nonumber\\
   I_4[\mathcal{A}]&=\int D_{\alpha_i} \int_0^1 dv~ \mathcal{A}(\alpha_{\bar q},\alpha_q,\alpha_g)
 \delta(\alpha_{\bar q}+ v \alpha_g-u_0),\nonumber\\
   I_5[\mathcal{A}]&=\int_0^1 du~ A(u)\delta'(u-u_0),\nonumber\\
 I_6[\mathcal{A}]&=\int_0^1 du~ A(u),
 \end{align}
 where $\mathcal{A}$ represents the corresponding photon DAs.

 \section{on-shell photon Distribution Amplitudes}\label{appendb}
In this appendix, we give the descriptions of the matrix elements of the form $\langle \gamma(q)\vel \bar{q}(x) \Gamma_i q(0) \ver 0\rangle$  
and $\langle \gamma(q)\vel \bar{q}(x) \Gamma_i G_{\mu\nu}q(0) \ver 0\rangle$ regarding the on-shell photon DAs along with the explicit expressions of the photon DAs entering into the matrix elements~\cite{Ball:2002ps}:
\begin{eqnarray}
\label{esbs14}
&&\langle \gamma(q) \vert  \bar q(x) \gamma_\mu q(0) \vert 0 \rangle
= e_q f_{3 \gamma} \left(\varepsilon_\mu - q_\mu \frac{\varepsilon
x}{q x} \right) \int_0^1 du e^{i \bar u q x} \psi^v(u),
\nonumber \\
&&\langle \gamma(q) \vert \bar q(x) \gamma_\mu \gamma_5 q(0) \vert 0
\rangle  = - \frac{1}{4} e_q f_{3 \gamma} \epsilon_{\mu \nu \alpha
\beta } \varepsilon^\nu q^\alpha x^\beta \int_0^1 du e^{i \bar u q
x} \psi^a(u),
\nonumber \\
&&\langle \gamma(q) \vert  \bar q(x) \sigma_{\mu \nu} q(0) \vert  0
\rangle  = -i e_q \langle \bar q q \rangle (\varepsilon_\mu q_\nu - \varepsilon_\nu
q_\mu) \int_0^1 du e^{i \bar u qx} \left(\chi \varphi_\gamma(u) +
\frac{x^2}{16} \mathbb{A}  (u) \right) \nonumber \\ 
&&-\frac{i}{2(qx)}  e_q \bar qq \left[x_\nu \left(\varepsilon_\mu - q_\mu
\frac{\varepsilon x}{qx}\right) - x_\mu \left(\varepsilon_\nu -
q_\nu \frac{\varepsilon x}{q x}\right) \right] \int_0^1 du e^{i \bar
u q x} h_\gamma(u),
\nonumber \\
&&\langle \gamma(q) | \bar q(x) g_s G_{\mu \nu} (v x) q(0) \vert 0
\rangle = -i e_q \langle \bar q q \rangle \left(\varepsilon_\mu q_\nu - \varepsilon_\nu
q_\mu \right) \int {\cal D}\alpha_i e^{i (\alpha_{\bar q} + v
\alpha_g) q x} {\cal S}(\alpha_i),
\nonumber \\
&&\langle \gamma(q) | \bar q(x) g_s \tilde G_{\mu \nu}(v
x) i \gamma_5  q(0) \vert 0 \rangle = -i e_q \langle \bar q q \rangle \left(\varepsilon_\mu q_\nu -
\varepsilon_\nu q_\mu \right) \int {\cal D}\alpha_i e^{i
(\alpha_{\bar q} + v \alpha_g) q x} \tilde {\cal S}(\alpha_i),
\nonumber \\
&&\langle \gamma(q) \vert \bar q(x) g_s \tilde G_{\mu \nu}(v x)
\gamma_\alpha \gamma_5 q(0) \vert 0 \rangle = e_q f_{3 \gamma}
q_\alpha (\varepsilon_\mu q_\nu - \varepsilon_\nu q_\mu) \int {\cal
D}\alpha_i e^{i (\alpha_{\bar q} + v \alpha_g) q x} {\cal
A}(\alpha_i),
\nonumber 
\end{eqnarray}
\begin{eqnarray}
&&\langle \gamma(q) \vert \bar q(x) g_s G_{\mu \nu}(v x) i
\gamma_\alpha q(0) \vert 0 \rangle = e_q f_{3 \gamma} q_\alpha
(\varepsilon_\mu q_\nu - \varepsilon_\nu q_\mu) \int {\cal
D}\alpha_i e^{i (\alpha_{\bar q} + v \alpha_g) q x} {\cal
V}(\alpha_i), \nonumber\\
&& \langle \gamma(q) \vert \bar q(x)
\sigma_{\alpha \beta} g_s G_{\mu \nu}(v x) q(0) \vert 0 \rangle  =
e_q \langle \bar q q \rangle \left\{
        \left[\left(\varepsilon_\mu - q_\mu \frac{\varepsilon x}{q x}\right)\left(g_{\alpha \nu} -
        \frac{1}{qx} (q_\alpha x_\nu + q_\nu x_\alpha)\right) \right. \right. q_\beta
\nonumber \\
 && -
         \left(\varepsilon_\mu - q_\mu \frac{\varepsilon x}{q x}\right)\left(g_{\beta \nu} -
        \frac{1}{qx} (q_\beta x_\nu + q_\nu x_\beta)\right) q_\alpha
-
         \left(\varepsilon_\nu - q_\nu \frac{\varepsilon x}{q x}\right)\left(g_{\alpha \mu} -
        \frac{1}{qx} (q_\alpha x_\mu + q_\mu x_\alpha)\right) q_\beta
\nonumber \\
 &&+
         \left. \left(\varepsilon_\nu - q_\nu \frac{\varepsilon x}{q.x}\right)\left( g_{\beta \mu} -
        \frac{1}{qx} (q_\beta x_\mu + q_\mu x_\beta)\right) q_\alpha \right]
   \int {\cal D}\alpha_i e^{i (\alpha_{\bar q} + v \alpha_g) qx} {\cal T}_1(\alpha_i)
\nonumber \\
 &&+
        \left[\left(\varepsilon_\alpha - q_\alpha \frac{\varepsilon x}{qx}\right)
        \left(g_{\mu \beta} - \frac{1}{qx}(q_\mu x_\beta + q_\beta x_\mu)\right) \right. q_\nu
\nonumber \\ &&-
         \left(\varepsilon_\alpha - q_\alpha \frac{\varepsilon x}{qx}\right)
        \left(g_{\nu \beta} - \frac{1}{qx}(q_\nu x_\beta + q_\beta x_\nu)\right)  q_\mu
\nonumber \\ && -
         \left(\varepsilon_\beta - q_\beta \frac{\varepsilon x}{qx}\right)
        \left(g_{\mu \alpha} - \frac{1}{qx}(q_\mu x_\alpha + q_\alpha x_\mu)\right) q_\nu
\nonumber \\ &&+
         \left. \left(\varepsilon_\beta - q_\beta \frac{\varepsilon x}{qx}\right)
        \left(g_{\nu \alpha} - \frac{1}{qx}(q_\nu x_\alpha + q_\alpha x_\nu) \right) q_\mu
        \right]      
    \int {\cal D} \alpha_i e^{i (\alpha_{\bar q} + v \alpha_g) qx} {\cal T}_2(\alpha_i)
\nonumber \\
&&+\frac{1}{qx} (q_\mu x_\nu - q_\nu x_\mu)
        (\varepsilon_\alpha q_\beta - \varepsilon_\beta q_\alpha)
    \int {\cal D} \alpha_i e^{i (\alpha_{\bar q} + v \alpha_g) qx} {\cal T}_3(\alpha_i)
\nonumber \\ &&+
        \left. \frac{1}{qx} (q_\alpha x_\beta - q_\beta x_\alpha)
        (\varepsilon_\mu q_\nu - \varepsilon_\nu q_\mu)
    \int {\cal D} \alpha_i e^{i (\alpha_{\bar q} + v \alpha_g) qx} {\cal T}_4(\alpha_i)
                        \right\},
\end{eqnarray}
where the ${\cal D} \alpha_i$ is defined as
\begin{eqnarray}
\label{nolabel05}
\int {\cal D} \alpha_i = \int_0^1 d \alpha_{\bar q} \int_0^1 d
\alpha_q \int_0^1 d \alpha_g \delta(1-\alpha_{\bar
q}-\alpha_q-\alpha_g)~.
\end{eqnarray}
Here, $\varphi_\gamma(u)$ denotes the leading twist-2 of the photon DA, $\psi^v(u)$,
$\psi^a(u)$, ${\cal A}(\alpha_i)$, and ${\cal V}(\alpha_i)$ are the twist-3 DAs, and
$h_\gamma(u)$, $\mathbb{A}(u)$, ${\cal S}(\alpha_i)$, ${\cal{\tilde S}}(\alpha_i)$, ${\cal T}_1(\alpha_i)$, ${\cal T}_2(\alpha_i)$, ${\cal T}_3(\alpha_i)$ 
and ${\cal T}_4(\alpha_i)$ are the twist-4 photon DAs. The explicit expressions of the on-shell photon DAs with different twists are
\begin{eqnarray}
\varphi_\gamma(u) &=& 6 u \bar u \left( 1 + \varphi_2(\mu)
C_2^{\frac{3}{2}}(u - \bar u) \right),
\nonumber \\
\psi^v(u) &=& 3 \left(3 (2 u - 1)^2 -1 \right)+\frac{3}{64} \left(15
w^V_\gamma - 5 w^A_\gamma\right)
                        \left(3 - 30 (2 u - 1)^2 + 35 (2 u -1)^4
                        \right),
\nonumber \\
\psi^a(u) &=& \left(1- (2 u -1)^2\right)\left(5 (2 u -1)^2 -1\right)
\frac{5}{2}
    \left(1 + \frac{9}{16} w^V_\gamma - \frac{3}{16} w^A_\gamma
    \right),
\nonumber \\
h_\gamma(u) &=& - 10 \left(1 + 2 \kappa^+\right) C_2^{\frac{1}{2}}(u
- \bar u),
\nonumber \\
\mathbb{A}(u) &=& 40 u^2 \bar u^2 \left(3 \kappa - \kappa^+
+1\right)  +
        8 (\zeta_2^+ - 3 \zeta_2) \left[u \bar u (2 + 13 u \bar u) \right.
\nonumber \\ && + \left.
                2 u^3 (10 -15 u + 6 u^2) \ln(u) + 2 \bar u^3 (10 - 15 \bar u + 6 \bar u^2)
        \ln(\bar u) \right],
\nonumber \\
{\cal A}(\alpha_i) &=& 360 \alpha_q \alpha_{\bar q} \alpha_g^2
        \left(1 + w^A_\gamma \frac{1}{2} (7 \alpha_g - 3)\right),
\nonumber \\
{\cal V}(\alpha_i) &=& 540 w^V_\gamma (\alpha_q - \alpha_{\bar q})
\alpha_q \alpha_{\bar q}
                \alpha_g^2,
\nonumber \\
{\cal T}_1(\alpha_i) &=& -120 (3 \zeta_2 + \zeta_2^+)(\alpha_{\bar
q} - \alpha_q)
        \alpha_{\bar q} \alpha_q \alpha_g,
\nonumber \\
{\cal T}_2(\alpha_i) &=& 30 \alpha_g^2 (\alpha_{\bar q} - \alpha_q)
    \left((\kappa - \kappa^+) + (\zeta_1 - \zeta_1^+)(1 - 2\alpha_g) +
    \zeta_2 (3 - 4 \alpha_g)\right),
\nonumber \\
{\cal T}_3(\alpha_i) &=& - 120 (3 \zeta_2 - \zeta_2^+)(\alpha_{\bar
q} -\alpha_q)
        \alpha_{\bar q} \alpha_q \alpha_g,
\nonumber \\
{\cal T}_4(\alpha_i) &=& 30 \alpha_g^2 (\alpha_{\bar q} - \alpha_q)
    \left((\kappa + \kappa^+) + (\zeta_1 + \zeta_1^+)(1 - 2\alpha_g) +
    \zeta_2 (3 - 4 \alpha_g)\right),\nonumber \\
{\cal S}(\alpha_i) &=& 30\alpha_g^2\{(\kappa +
\kappa^+)(1-\alpha_g)+(\zeta_1 + \zeta_1^+)(1 - \alpha_g)(1 -
2\alpha_g)\nonumber +\zeta_2[3 (\alpha_{\bar q} - \alpha_q)^2-\alpha_g(1 - \alpha_g)]\},\nonumber \\
\tilde {\cal S}(\alpha_i) &=&-30\alpha_g^2\{(\kappa -\kappa^+)(1-\alpha_g)+(\zeta_1 - \zeta_1^+)(1 - \alpha_g)(1 -
2\alpha_g)
+\zeta_2 [3 (\alpha_{\bar q} -\alpha_q)^2-\alpha_g(1 - \alpha_g)]\}.
\end{eqnarray}
The numerical values of the constants in the above wave functions are given as $\varphi_2(1~GeV) = 0$, 
$w^V_\gamma = 3.8 \pm 1.8$, $w^A_\gamma = -2.1 \pm 1.0$, $\kappa = 0.2$, $\kappa^+ = 0$, $\zeta_1 = 0.4$, and $\zeta_2 = 0.3$.

\end{widetext}

\bibliographystyle{elsarticle-num}
\bibliography{Zc-diquark.bib}

\end{document}